\documentclass[aps,prl,twocolumn,superscriptaddress,showpacs,amsmath,amssymb,footinbib]{revtex4-1}
\usepackage{mathtools}
\usepackage{amsfonts}
\usepackage{amssymb}
\usepackage{graphicx}
\usepackage{color}

\usepackage{bm}
\usepackage{braket}
\usepackage{gensymb}

\usepackage{hyperref}

\newcommand{\calT}{\mathcal{T}}

\DeclareMathOperator{\Tr}{Tr}

\begin{document}

\title{Quantum Quench and $f$-Sum Rules on Linear and Non-linear Conductivities}

\author{Masaki Oshikawa}
\affiliation{Institute for Solid State Physics, University of Tokyo, Kashiwa 277-8581, Japan.}
\affiliation{Kavli Institute for the Physics and Mathematics of the Universe (WPI), University of Tokyo, Kashiwa 277-8583, Japan}

\author{Haruki Watanabe}
\affiliation{Department of Applied Physics, University of Tokyo,
Tokyo 113-8656, Japan.}

\date{\today}

\begin{abstract}
Considering a quench process in which an electric field pulse is applied
to the system,  ``$f$-sum rule'' for the conductivity
for general quantum many-particle systems is derived.
It is furthermore extended to an infinite series of sum rules,
applicable to the nonlinear  conductivity at every order.
\end{abstract}

\maketitle

\clearpage

\textit{Introduction}.---
Understanding of dynamical responses of a quantum many-body system is
not only theoretically interesting but is also essential for bridging
theory and experiment, as many experiments measure dynamical responses.
Linear responses have been best understood, thanks to the general
framework of linear response theory~\cite{Kubo-StatPhysII}.
Many experiments can be actually well
described in terms of linear responses.
On the other hand, there is a renewed strong interest in nonlinear
responses recently,
thanks to new theoretical ideas, powerful numerical methods, and
developments in experimental techniques such as powerful laser sources
which enable us to probe highly nonlinear responses.  For example,
``shift current'', which is a DC current induced by AC electric field as
a higher order effect, has been studied
vigorously~\cite{SipeShkrebtii2000,YoungRappe2012,MorimotoNagaosa2016,MorimotoNagaosa2018,Yang-shift2017}.

Yet, theoretical computations of dynamical responses are generally challenging,
often even for linear responses and more so for nonlinear ones.
Therefore it is useful to obtain general constraints on dynamical
responses, including their relations to \emph{static} quantities which are
easier to calculate.
The ``$f$-sum rule'' of the linear electric conductivity is a typical
and well-known example of such constraints~\cite{Pines-EE_Solids}.
For simplicity, here let us consider the uniform component
of the linear AC conductivity, which is defined as
\begin{equation}
 j_i(\omega) = \sigma_{i}^j(\omega) E_j(\omega),
\end{equation}
where $j_i(\omega)=j_i(-\omega)^*$ is the uniform part ($\bm{q}=0$ Fourier component) of the
current, $E_j(\omega)=E_j(-\omega)^*$  is the uniform electric field, and $\omega$
is the angular frequency. The $f$-sum rule is a constraint on the frequency integral $\int_{-\infty}^{\infty} \frac{d\omega}{2\pi} \; \sigma_{i}^j(\omega)$.

In condensed matter physics, the Hamiltonian of the following form
is often considered:
\begin{equation}
 \hat{H} = \hat{K} + \hat{I},
\label{eq.H_in_KI}
\end{equation}
where $\hat{K}$ is the kinetic energy (including the chemical potential term)
which is bilinear in particle creation/annihilation operators,
and $\hat{I}$ is the density-density interaction energy.  For the standard kinetic term in nonrelativistic quantum mechanics in the continuum
\begin{align}
 \hat{K} = \int d\bm{r} \; \hat{\psi}^\dagger(\bm{r})
 \left( - \frac{\bm{\nabla}^2}{2m} - \mu \right)
\hat{\psi}(\bm{r}),
\label{eq.K_NR}
\end{align}
the original form of the $f$-sum rule is known as
\begin{equation}
  \int_{-\infty}^{\infty} \frac{d\omega}{2\pi}\; \sigma_{i}^j(\omega)
   = \delta_{ij}\frac{\rho}{2m}.
\label{eq.fsum_NR}
\end{equation}
The right-hand side is determined by the electron mass $m$ and the
electron density $\rho$, and is a completely static quantity.

For more general models of the form~\eqref{eq.H_in_KI}, the
$f$-sum rule still holds although with a modified right hand
side~\cite{Bari-fsum,Sadataka-fsum,Izuyama1973,Maldague1977,BaeriswylCarmeloLuther1986,ShastrySutherland-twisted,RigolShastry-Drude,LimtragoolPhillips2017,Hazra2018}.
Namely, for a Hamiltonian of the form~\eqref{eq.H_in_KI}
where the kinetic term is
\begin{align}
\hat{K} &= \int \frac{d\bm{p}}{(2\pi)^d}
\hat{\psi}^\dagger(\bm{p}) \epsilon(\bm{p}) \hat{\psi}(\bm{p}) 
\end{align}
in the momentum representation, the $f$-sum rule reads
\begin{align}
 \int_{-\infty}^{\infty} \frac{d\omega}{2\pi}\; \sigma_{i}^j(\omega)
&= \frac{1}{2}
\int \frac{d\bm{p}}{(2\pi)^d}
\langle
\hat{\psi}^\dagger(\bm{p})
\frac{\partial^2 \epsilon(\bm{p})}{\partial p_i\partial p_j} \hat{\psi}(\bm{p})
\rangle.
\label{eq.fsum_generalK}
\end{align}

In this paper, we relate the $f$-sum rule to a quantum quench process.
This picture naturally leads to more general $f$-sum rules than
the form~\eqref{eq.fsum_generalK} that have been discussed in the literature.
In particular, we derive an infinite
series of ``$f$-sum rules'' for nonlinear conductivities.

\begin{figure}
\begin{center}
\includegraphics[width=0.99\columnwidth]{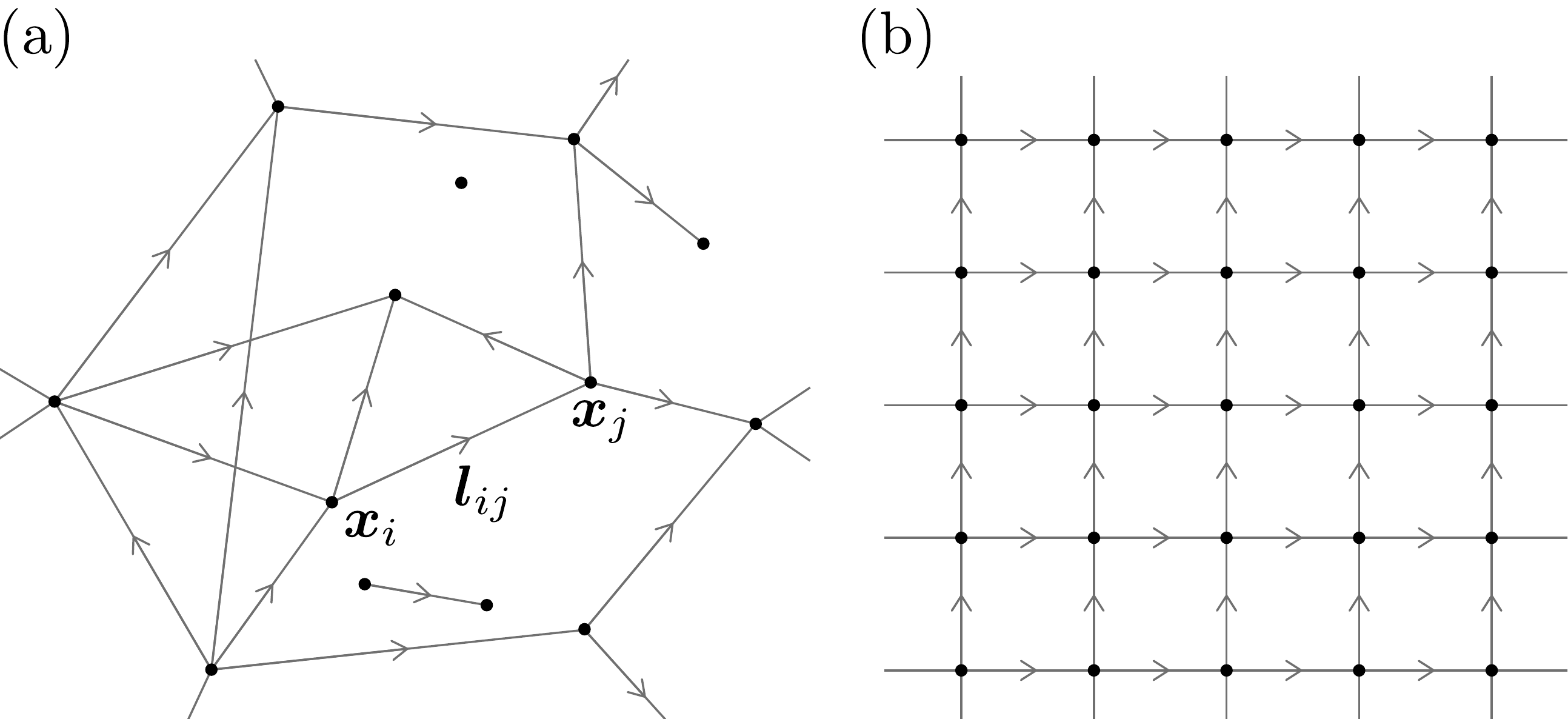}
\caption{\label{fig} (a) A general lattice. We assume that there are no loops (arrows that connects a point to itself), no multiple links to a single pair of points, or no bidirected arrows. (Such a graph is called ``simple'' and ``oriented'' in graph theory.)  Also, not all points have to be connected to each other. (b) Square lattice.}
\end{center}
\end{figure}

\textit{Setup and Result}.---
We consider a general system of many quantum particles defined on a $d$-dimensional lattice.
Let $P$ be the set of lattice points and $L$ be the set of directed links (arrows)  connecting a pair of lattice points as shown in Fig.~\ref{fig} (a).  We do not require any spatial symmetry such as the translation invariance or the inversion symmetry. The system size and the boundary condition can be chosen arbitrary.

The Hamiltonian $\hat{H}(t)$ of the system is written in terms of creation and annihilation operators $\hat{c}_{\bm{x}\alpha}^\dagger$, $\hat{c}_{\bm{x}\alpha}$ ($\alpha$ labels the internal degrees of freedom) defined on each point $\bm{x}\in P$ and a U(1) vector potential $A_{\bm{l}}(t)$ defined on each link $\bm{l}\in L$, while the scalar potential is set to be $0$.  The vector potential is introduced as an external field and its time dependence describes the local electric field~\footnote{To avoid negative signs, we use the sign convention of $A_{\bm{l}}(t)$ opposite to the standard definition.}\footnote{In general, $A_{\bm{l}}(t)$ can also produce a local magnetic field. However, the effect of the magnetic field on the induced current is suppressed by a factor of $T$ (duration of the time evolution)
and is neglected in the quench limit $T\rightarrow0$ considered in this paper.}
\begin{equation}
E_{\bm{l}'}(t)\equiv\frac{dA_{\bm{l}'}(t)}{dt}.\label{defE}
\end{equation}
We assume that the Hamiltonian is invariant under the local U(1) transformation $\hat{c}_{\bm{x}_i\alpha}\rightarrow\hat{c}_{\bm{x}_i\alpha}e^{i\theta_{\bm{x}_i}}$ and $A_{\bm{l}_{ij}}(t)\rightarrow A_{\bm{l}_{ij}}(t)-\theta_{\bm{x}_{j}}+\theta_{\bm{x}_{i}}$ where the link $\bm{l}_{ij}\in L$ goes from $\bm{x}_i\in P$ to $\bm{x}_j\in P$.  This enables us to define the conserved current density 
\begin{equation}
\hat{j}_{\bm{l}}(t)\equiv\frac{\partial \hat{H}(t)}{\partial A_{\bm{l}}(t)} \label{defj}
\end{equation}
at every link. 
We allow any number of creation and annihilation operators to appear in a single term in the Hamiltonian, representing correlated hopping,
pair hopping, ring exchange, and so on. We assume that all the hoppings and interactions are short-ranged and that the Hamiltonian depends on $t$ only through $A_{\bm{l}}(t)$~\footnote{In fact, all of our results hold even when the Hamiltonian has time dependence in addition to those originating from $A_{\bm{l}}(t)$, as far as such additional dependence is smooth, i.e., not of the quench type.}.

Suppose that the system is described by a density operator $\hat{\rho}(0)$ at $t=0$.  
The evolution of the system is given by the time-evolution operator $\hat{S}(t)$ defined by
\begin{align}
\frac{d\hat{S}(t)}{dt}=-i \hat{H}(t)\hat{S}(t),\quad \hat{S}(0)=1.\label{dS}
\end{align}
The expectation value of any operator $\hat{O}(t)$ at time $t$ is then given by 
\begin{equation}
\langle \hat{O}(t) \rangle_t \equiv\Tr[\hat{O}(t)\hat{S}(t)\hat{\rho}(0)\hat{S}(t)^\dagger].\label{rhot}
\end{equation}

The linear and nonlinear conductivities in real space and time are defined as the response of the the local current density as a result of the applied electric field:
\begin{align}
\langle\hat{j}_{\bm{l}}(t)\rangle_t&=\langle\hat{j}_{\bm{l}}(0)\rangle_0\notag\\
&+\sum_{\bm{m}\neq\bm{0}}\int_0^td\bm{t}'\,\sigma_{\bm{l}}^{\bm{m}}(t,\bm{t}')\prod_{\bm{l}'\in L}\frac{1}{m_{\bm{l}'}!}\prod_{i=1}^{m_{\bm{l}'}}E_{\bm{l}'}(t_{\bm{l}'i}).
\label{defsigma}
\end{align}
Here, $\bm{t}'$ is the collection of $t_{\bm{l}'i}$ and $\int_0^td\bm{t}'\equiv\prod_{\bm{l}'\in L}\prod_{i=1}^{m_{\bm{l}'}}\int_0^tdt_{\bm{l}'i}$ is the convolution integral.  Also, $\bm{m}$ is the collection of $m_{\bm{l}}\geq0$ and 
\begin{equation}
N\equiv\sum_{\bm{l}}m_{\bm{l}}
\end{equation}
represents the order of responses.  Namely, $\sigma_{\bm{l}}^{\bm{m}}(t,\bm{t}')$ with $N=1$ represents the (spatially-resolved) linear conductivity, whereas the case with $N\geq2$ corresponds to non-linear conductivities.

Our main result of this work is the following constraint on instantaneous conductivities for any $\bm{m}\neq\bm{0}$:
\begin{align}
&\sigma_{\bm{l}}^{\bm{m}}(0,\bm{0})=\langle\hat{H}^{\bm{m}|_{m_{\bm{l}}\rightarrow m_{\bm{l}}+1}}\rangle_0,\label{main}\\
&\hat{H}^{\bm{m}}\equiv\prod_{\bm{l}\in L}\frac{\partial^{m_{\bm{l}}}}{\partial A_{\bm{l}}(t)^{m_{\bm{l}}}}\hat{H}(t)\Big|_{t=0}.
\end{align}
In Eq.~\eqref{main}, $\sigma_{\bm{l}}^{\bm{m}}(0,\bm{0})$ is the shorthand for 
\begin{equation}
\lim_{t\rightarrow+0}\lim_{\bm{t}'\rightarrow+\bm{0}}\sigma_{\bm{l}}^{\bm{m}}(t,\bm{t}')
\end{equation}
and $m_{\bm{l}}\rightarrow m_{\bm{l}}+1$ means replacing $m_{\bm{l}}$ in $\bm{m}$ with $m_{\bm{l}}+1$. 
This spatially-resolved formula implies, among other things, that the instantaneous response for short-range hopping models vanishes
when $\bm{l}$ and $\bm{l}'$ are sufficiently apart. This is consistent with the intuition, and also with the
Lieb-Robinson bound~\cite{LiebRobinson-bound}.  

The constraint~\eqref{main} is valid on the instantaneous response in an
arbitrary initial state $\hat{\rho}(0)$.
A natural choice of $\hat{\rho}(0)$ would be an equilibrium density
matrix (Gibbs state) at a certain temperature, but
$\hat{\rho}(0)$ can also be chosen to represent a non-equilibrium state~\cite{ShimizuYuge1,ShimizuYuge2}, especially
a non-equilibrium steady state for which the response function
would still be time-translation invariant~\footnote{
we thank Kazuaki Takasan for suggesting potential applications
to non-equilibrium steady states.}.
In a steady state including the equilibrium,
$\sigma_{\bm{l}}^{\bm{m}}(t,\bm{t}')$ depends only on the
time differences $\Delta t_{\bm{l}'i} = t - t_{\bm{l}'i}$.
In such a case, it is common to work in the frequency space
after a Fourier transformation on $\Delta t_{\bm{l}'i}$. Then the left hand side of Eq.~\eqref{main} reads
\begin{align}
\sigma_{\bm{l}}^{\bm{m}}(0,\bm{0})=2^N
\int_{-\infty}^{\infty} d \bm{\omega} \; \sigma_{\bm{l}}^{\bm{m}}(\bm{\omega}),
\label{eq.fsum_n}
\end{align}
where $\bm{\omega}$ is the collection of $\omega_{\bm{l}'i}$ and $\int_{-\infty}^{\infty} d \bm{\omega}\equiv\prod_{\bm{l}'\in L}\prod_{i=1}^{m_{\bm{l}'}}\int_{-\infty}^{\infty}\frac{d\omega_{\bm{l}'i}}{2\pi}$.  The factor $2^{N}$ originates from the discontinuity of $\sigma_{\bm{l}}^{\bm{m}}$ around $\Delta t_{\bm{l}'i}=0$.  Eqs.~\eqref{main} and \eqref{eq.fsum_n} give general, position-dependent $f$-sum rules in terms of the
frequency integral.

\textit{Example: uniform response on square lattice}.---
To illustrate implications of our result in the simplest setting, let us take the 2D square lattice and discuss the response of the averaged current density toward a uniform electric field.  We assign the common value of the vector potential $A_x(t)$ ($A_y(t)$) to the horizontal (vertical) links in Fig.~\ref{fig} (b). In this case, the averaged current density is defined by
\begin{equation}
\hat{j}_{i}(t)=\frac{1}{V}\frac{\partial\hat{H}(t)}{\partial A_{i}(t)} \quad (i=x,y)
\end{equation}
($V$ is the volume of the system) and the definition of conductivities in Eq.~\eqref{defsigma} is simplified to
\begin{align}
&\langle\hat{j}_{i}(t)\rangle_t=\langle\hat{j}_{i}(0)\rangle_0\notag\\
&+\int_0^tdt_1\left[\sigma_{i}^{(1,0)}(t,t_1)E_x(t_1)+\sigma_{i}^{(0,1)}(t,t_1)E_y(t_1)\right]\notag\\
&+\int_0^tdt_1\int_0^tdt_2\Big[\frac{1}{2}\sigma_{i}^{(2,0)}(t,t_1,t_2)E_x(t_1)E_x(t_2)\notag\\
&\quad\quad\quad\quad\quad+\sigma_{i}^{(1,1)}(t,t_1,t_2)E_x(t_1)E_y(t_2)\notag\\
&\quad\quad\quad\quad+\frac{1}{2}\sigma_{i}^{(0,2)}(t,t_1,t_2)E_y(t_1)E_y(t_2)\Big]+\ldots.
\label{defsigma2}
\end{align}

Our result \eqref{main} reproduces the well-known $f$-sum rule
on the linear conductivity~\footnote{
In Ref.~\protect\cite{Izuyama1973},
a breakdown of the sum rule for the uniform $q=0$ component was
discussed. However the issue is presumably related to the subtlety of the
Drude peak. In the perfectly periodic system studied in the present
paper,  at least in a finite-size system,
the sum rule is exactly satisfied by including the possible
Drude peak in the integral,
as implied by the argument presented in the main text.
}:
\begin{align}
&\sigma_{x}^{(1,0)}(0,0)=\frac{1}{2V}\langle\frac{\partial^2\hat{H}}{\partial A_x^2}\rangle_0\Big|_{t=0},\\
&\sigma_{x}^{(0,1)}(0,0)=\sigma_{y}^{(1,0)}(0,0)=\frac{1}{2V}\langle\frac{\partial^2\hat{H}}{\partial A_x\partial A_y}\rangle_0\Big|_{t=0},\label{symmetric}\\
&\sigma_{y}^{(0,1)}(0,0)=\frac{1}{2V}\langle\frac{\partial^2\hat{H}}{\partial A_y^2}\rangle_0\Big|_{t=0}.
\end{align}
Note that Eq.~\eqref{symmetric} is derived without assuming any spatial symmetry.  

We also have an infinite series of $f$-sum rules on nonlinear conductivities. For example, relations at the quadratic order read
\begin{align}
&\sigma_{x}^{(2,0)}(0,0,0)=\frac{1}{4V}\langle\frac{\partial^3\hat{H}}{\partial A_x^3}\rangle_0\Big|_{t=0},\\
&\sigma_{x}^{(1,1)}(0,0,0)=\sigma_{y}^{(2,0)}(0,0,0)=\frac{1}{4V}\langle\frac{\partial^3\hat{H}}{\partial A_x^2\partial A_y}\rangle_0\Big|_{t=0},\\
&\sigma_{x}^{(0,2)}(0,0,0)=\sigma_{y}^{(1,1)}(0,0,0)=\frac{1}{4V}\langle\frac{\partial^3\hat{H}}{\partial A_x\partial A_y^2}\rangle_0\Big|_{t=0},\\
&\sigma_{y}^{(0,2)}(0,0,0)=\frac{1}{4V}\langle\frac{\partial^3\hat{H}}{\partial A_y^3}\rangle_0\Big|_{t=0}.
\end{align}

When there are only nearest neighbor hoppings on the square lattice,
the right-hand side of higher-order $f$-sum rules are reduced to the one for the linear response and the persistent current. For example, we have
\begin{align}
\sigma_x^{(n,0)}(0,\bm{0}) = 
\begin{dcases}
 \frac{ (-1)^{(n+1)/2}}{2^nV} \langle \frac{\partial^2 \hat{H}}{\partial A_x^2} \rangle_0\Big|_{t=0} &
 \mbox{($n$: odd)},
\\
  (-1)^{n/2} 2^{-n}\langle \hat{j}_x(0) \rangle_0  & \mbox{($n$: even)}.
\end{dcases}
\end{align}

\textit{Derivation of the Main Result}.---To demonstrate our result~\eqref{main}, we choose the vector potential on the link $\bm{l}$ to be
\begin{equation}
A_{\bm{l}}(t)=f_{\bm{l}}(t/T)\mathcal{A}_{\bm{l}},
\end{equation}
where $\mathcal{A}_{\bm{l}}$ is a constant and $f_{\bm{l}}(\tau)$ is a smooth function satisfying $f_{\bm{l}}(\tau)=0$ for $\tau<0$ and $f_{\bm{l}}(\tau)=1$ for $\tau>1$.

We start with verifying
\begin{equation}
\frac{d}{dt} \langle \hat{H}(t) \rangle_t =\langle\frac{d\hat{H}(t)}{dt} \rangle_t=\sum_{\bm{l}\in L}E_{\bm{l}}(t)\langle \hat{j}_{\bm{l}}(t) \rangle_t
\label{dtdH}
\end{equation}
by combining Eqs.~\eqref{defE}--\eqref{rhot}. Integrating this equation over $t\in[0,T]$, we get
\begin{align}
\langle\hat{H}(T)\rangle_T-\langle\hat{H}(0)\rangle_0=\sum_{\bm{l}\in L}\int_0^TdtE_{\bm{l}}(t)\langle\hat{j}_{\bm{l}}(t)\rangle_t.
\label{eq.integrated2}
\end{align}
A relation similar to \eqref{eq.integrated2} for the uniform current and electric field was
used in Ref.~\cite{Oshikawa-Drude,*Oshikawa-Drude-Erratum}.
There, the limit $T \to \infty$ of the adiabatic
flux insertion was considered, in order to discuss the Drude weight
at zero temperature $\beta \to \infty$. The adiabatic insertion leads to the famous Kohn formula for
the Drude weight~\cite{Kohn1964}.
However, the formula~\eqref{eq.integrated2} is valid for general
$T$ and for any initial state.
Here we consider the opposite limit, that is the
limit of very quick insertion of flux: $T \to 0$. This can be regarded as an example of quantum quench
(sudden switching of the vector potential).
In this limit, the state cannot follow the change of the Hamiltonian, and
``the sudden approximation $\hat{S}(t)=1$'' becomes exact~\footnote{The fact that $\hat{S}(t)\rightarrow1$ in the $T\to0$ limit might sound puzzling, since the applied electric field may still give a non-zero impulse to the system even in the quench limit. This is not a contradiction because the impulse is \emph{not} described by $\hat{S}(t)$ but by the (large) gauge transformation that brings $\hat{H}(T)$ back to $\hat{H}(0)$, although here we do not perform such a transformation.}.  This can be most easily seen by the formula ($\calT$ denotes the time-ordering)
\begin{align}
\hat{S}(t)=\mathcal{T} e^{ -i T\int_0^{t/T} d\tau\hat{H}[f_{\bm{l}}(\tau)\mathcal{A}_{\bm{l}}]}.
\end{align}
Because of the prefactor $T$ in the exponent, $\hat{S}(t)\to 1$ in the limit of $T\to0$ for any $0\leq t\leq T$.  In this limit, only the ``diamagnetic" contributions survive in the current response.

In the following, we expand each side of Eq.~\eqref{eq.integrated2} into the power series of $\mathcal{A}_{\bm{l}}$ in the quench limit $T\to0$.  On the one hand, the left-hand side of Eq.~\eqref{eq.integrated2} is reduced to $\langle\hat{H}(T)-\hat{H}(0)\rangle_0$, which admits the Taylor expansion
\begin{align}
\langle\hat{H}(T)-\hat{H}(0)\rangle_0=\sum_{\bm{m}\neq\bm{0}}\langle\hat{H}^{\bm{m}}\rangle_0\prod_{\bm{l}'\in L}\frac{1}{m_{\bm{l}'}!}\mathcal{A}_{\bm{l}'}^{m_{\bm{l}'}}.\label{LEFT}
\end{align}
On the other hand, we approximate $\sigma_{\bm{l}}^{\bm{m}}(t,\bm{t}')$ in Eq.~\eqref{defsigma} by $\sigma_{\bm{l}}^{\bm{m}}(0,\bm{0})$ assuming that $T$ is small enough.  We can then easily perform the $\int_0^td\bm{t}'$ integral and get
\begin{align}
&\langle\hat{j}_{\bm{l}}(t)\rangle_t=\langle\hat{j}_{\bm{l}}(0)\rangle_0\notag\\
&\quad+\sum_{\bm{m}\neq\bm{0}}\sigma_{\bm{l}}^{\bm{m}}(0,\bm{0})\prod_{\bm{l}'\in L}\frac{1}{m_{\bm{l}'}!}f_{\bm{l}'}(t/T)^{m_{\bm{l}'}}\mathcal{A}_{\bm{l}'}^{m_{\bm{l}'}}.
\end{align}
Thus the right-hand side of Eq.~\eqref{eq.integrated2} becomes
\begin{align}
&\sum_{\bm{l}\in L}\int_0^TdtE_{\bm{l}}(t)\langle\hat{j}_{\bm{l}}(t)\rangle_t=\sum_{\bm{l}\in L}\mathcal{A}_{\bm{l}}\langle\hat{j}_{\bm{l}}(0)\rangle_0\notag\\
&+\sum_{\bm{m}|\sum_{\bm{l}} m_{\bm{l}}\geq2}\left[\sum_{\bm{l}\in L}\sigma_{{\bm{l}}}^{\bm{m}|_{m_{\bm{l}}\rightarrow m_{\bm{l}}-1}}(0,\bm{0}) I_{\bm{l}}^{\bm{m}}\right]\prod_{\bm{l}'\in L}\frac{1}{m_{\bm{l}'}!}\mathcal{A}_{\bm{l}'}^{m_{\bm{l}'}},\label{RIGHT}
\end{align}
where 
\begin{align}
&I_{\bm{l}}^{\bm{m}}\equiv\int_0^1d\tau\partial_\tau[f_{\bm{l}}(\tau)^{m_{\bm{l}}}]\prod_{\bm{l}'\neq \bm{l}}f_{\bm{l}'}(\tau)^{m_{\bm{l}'}}.
\end{align}
When $m_{\bm{l}}=0$, $\sigma_{{\bm{l}}}^{\bm{m}|_{m_{\bm{l}}\rightarrow m_{\bm{l}}-1}}(0,\bm{0})$ is ill-defined but in this case $I_{\bm{l}}^{\bm{m}}$ vanishes and Eq~\eqref{RIGHT} still holds.

Matching the coefficient of $\prod_{\bm{l}'\in L}\frac{1}{m_{\bm{l}'}!}\mathcal{A}_{\bm{l}'}^{m_{\bm{l}'}}$ in Eqs.~\eqref{LEFT} and \eqref{RIGHT}, we find
\begin{equation}
\langle\hat{H}^{\bm{m}}\rangle_0=\sum_{\bm{l}\in L}\sigma_{{\bm{l}}}^{\bm{m}|_{m_{\bm{l}}\rightarrow m_{\bm{l}}-1}}(0,\bm{0}) I_{\bm{l}}^{\bm{m}}
\label{LR}
\end{equation}
for $N\geq2$.  Note that the integral $I_{\bm{l}}^{\bm{m}}$ depends on the specific choice of the function $f_{\bm{l}}(\tau)$. To avoid contradiction, we demand the invariance of the right-hand side of Eq.~\eqref{LR} under an arbitrary variation $\delta f_{\bm{l}}(\tau)$ with $\delta f_{\bm{l}}(0)=\delta f_{\bm{l}}(1)=0$. It implies
\begin{equation}
\sigma_{{\bm{l}}}^{\bm{m}|_{m_{\bm{l}}\rightarrow m_{\bm{l}}-1}}(0,\bm{0}) =\sigma_{{\bm{l}'}}^{\bm{m}|_{m_{\bm{l}'}\rightarrow m_{\bm{l}'}-1}}(0,\bm{0}) 
\end{equation}
for any pairs of $\bm{l}$ and $\bm{l}'$ with $m_{\bm{l}}\geq1$ and $m_{\bm{l}}'\geq1$. Plugging this relation back to Eq.~\eqref{LR} and using $\sum_{\bm{l}\in L}I_{\bm{l}}^{\bm{m}}=\int_0^1d\tau\partial_\tau[\prod_{\bm{l}'\in L}f_{\bm{l}'}(\tau)^{m_{\bm{l}'}}]=1$, we recover our main result in Eq.~\eqref{main}.

\textit{Discussion}.---
In this work, we obtained an infinite series of sum rules
on the nonlinear conductivities, although
just one sum rule for $\sigma_{\bm{l}}^{\bm{m}}$,
which has multiple arguments, was found. We stress that the present approach is quite general and not limited to the Hamiltonians~\eqref{eq.H_in_KI}.
It can be also naturally understood that the density-density
interactions do not appear explicitly in the sum rule:
any term in Hamiltonian which does not couple to the gauge field does not contribute to $H^{\bm{m}}$.  

While we used lattice models in our derivation,
essentially the same argument applies to continuum models as well.  For the particular case of the nonrelativistic quantum mechanical
Hamiltonian~\eqref{eq.H_in_KI} with~\eqref{eq.K_NR},
the right-hand side of the main result, Eq.~\eqref{main}, vanishes
for nonlinear conductivities.
Although this is rather remarkable, this does not imply the absence
of a nonlinear current response to the electric field.
Eq.~\eqref{main} just represents the instantaneous response,
and even when it vanishes, the response can be non-vanishing at a later time.
In the frequency representation~\eqref{eq.fsum_n},
the vanishment of Eq.~\eqref{main}
implies that any positive part of $\sigma_{i}^{\bm{m}}(\bm{\omega})$
must be compensated by a negative part.

The present result is one of rather
few general constraints on conductivities, especially non-linear ones.  The sum rules can be used to check various approximations or numerical
calculations, and might give a guiding principle on designing 
systems with  desired transport properties.
We hope that the present result will help developing theory
of linear and nonlinear dynamical responses of
quantum many-body systems in the future.

\begin{acknowledgments}
This work is initiated while M.~O. was
participating in the Harvard CMSA Program on
\textit{Topological Aspects of Condensed Matter}.
He thanks Yuan-Ming Lu, Ying Ran, and Xu Yang, for the discussions
during the Program which led to the present work.
A part of the work by M.~O. was also performed at
the Aspen Center for Physics, which is supported
by National Science Foundation Grant PHY-1607611.
We are grateful to Kazuaki Takasan, Takahiro Morimoto, Naoto Nagaosa,
Marcos Rigol, and Sriram Shastry
for very useful comments on the early version of the draft.
We also acknowledge useful discussions, including collaborations
on related earlier projects, with
Yoshiki Fukusumi, Shunsuke C. Furuya, Ryohei Kobayashi,
Gr\'{e}goire Misguich, Yuya Nakagawa, and Masaaki Nakamura.
The work of M.O. was supported in part by MEXT/JSPS KAKENHI
Grant Nos. JP19H01808 and JP17H06462.
The work of H.W. is supported by JST PRESTO Grant No. JPMJPR18LA.
\end{acknowledgments}

\bibliography{bibs}

\end{document}